\documentstyle[epsfig,12pt]{elsart} 
\begin{document}
\newcommand{\kp}{K^+}
\newcommand{\gk}{\vec{\gamma}\vec{k}}
\newcommand{\gE}{\gamma_0 E_k}
\newcommand{\ppl}{\vec{p}}
\newcommand{\bcm}{\vec{b}^{\star}}
\newcommand{\becm}{\vec{\beta}^{\star}}
\newcommand{\bepl}{\vec{\beta}}
\newcommand{\rcm}{\vec{r}^{\star}}
\newcommand{\rpl}{\vec{r}}
\newcommand{\A}{{$\mathcal A$}}
\newcommand{\wpk}{ \omega_{p-k}}
\newcommand{\Journal}[4]{ #1 {\bf #2} (#4) #3}
\newcommand{\NPA}{Nucl.\ Phys.\ A}
\newcommand{\PLB}{Phys.\ Lett.\ B}
\newcommand{\PRC}{Phys.\ Rev.\ C}
\newcommand{\ZPC}{Z.\ Phys.\ C}
\newcommand{\be}{\begin{equation}}
\newcommand{\ee}{\end{equation}}
\begin{frontmatter}

\title{Formation of hypernuclei in high energy reactions within a 
covariant transport model}

\author{T. Gaitanos}
\author{H. Lenske}
\author{U. Mosel}

\address{Institut f\"ur Theoretische Physik,
Universit\"at Giessen, D-35392, Giessen, Germany}
\address{email: Theodoros.Gaitanos@theo.physik.uni-giessen.de}
\begin{abstract}
We investigate the formation of fragments with strangeness degrees of 
freedom in proton- and heavy-ion-induced reactions at high relativistic 
energies. The model used is a combination of a dynamical transport model 
and a statistical approach of fragment formation. 
We discuss in detail the applicability and limitations 
of such a hybrid model by comparing data on spectator fragmentation at 
relativistic $SIS/GSI$-energies. The theoretical 
results are analyzed in terms of spectator fragmentation with 
strangeness degrees of freedom such as the production of 
single-$\Lambda-{}^{3,4,5}He$ hypernuclei. We provide theoretical 
estimates on the spectra and on inclusive cross sections of light 
hypernuclei, which could be helpful for future 
experiments on hypernuclear physics at the new GSI- and J-PARC-facilities.
\end{abstract}
\end{frontmatter}

\date{\today}

\section{Introduction}
Hypernuclear physics opens the unique possibility 
to investigate the properties of the hyperon-nucleon ($YN$) and the 
hyperon-hyperon ($YY$) interaction (a historical overview can be 
found in Ref. \cite{HypFirst1}). Information on the strangeness sector 
of the hadronic equation of state is essential for nuclear 
astrophysics, e.g., for the properties of neutron stars \cite{NeutronStars}, 
and in exploring exotic states of finite nuclei, e.g., neutron-rich 
strange nuclei and exotic di-baryonic systems \cite{HypNS}. 

Hypernuclear production in reactions between heavy nuclei was first 
theoretically proposed by Kerman and Weiss \cite{HypFirst2}. These authors 
found high energy reactions as the best possibility to create 
exotic finite nuclear systems with finite strangeness. Since then this field 
of research has been extended and attracted mainly theoretical interest, 
see, e.g., Refs. \cite{Wakai1,Wakai2,WakaiRest}. 

The experimental situation has so far been rather poor due to the short 
life time of hypernuclei \cite{HypDecay}, which impedes 
the detection of hypernuclei. However, the life time of hypernuclei 
seen in the laboratory is considerably enhanced in relativistic collisions 
due to relativistic effects. Hypernuclear production in proton-induced 
reactions has been experimentally studied also at COSY \cite{cosy} and 
the comparison with theoretical predictions based on transport equations 
of the Boltzmann type has been successfully performed \cite{cass}. 
More recently, concrete experimental proposals with high energy heavy-ion 
and proton beams at GSI (Darmstadt, Germany) and J-PARC (Japan), respectively, 
have been suggested by T. Saito \cite{HypHI,JPARC}. 

The production of hypernuclei by relativistic protons has been 
investigated before by non-relativistic transport theory \cite{cass} and 
fully quantum mechanically for coherent reactions \cite{shyam}. In this 
Letter we use a covariant formulation of transport theory. The approach 
is first applied to the production of hypernuclei in 
relativistic heavy ion collisions, as planned at GSI and FAIR. 
Secondly, we also consider proton-induced reactions, but in the 
J-PARC energy regime of 50 GeV. 
The initial and intermediate non-equilibrium stages are described 
by covariant transport theory based on Quantum-Hadro-Dynamics (QHD) 
\cite{QHD}. The formation of fragments in the exit 
channels is treated in the statistical multifragmentation model (SMM) 
\cite{botvina}, including different models of fragment formation (apart from 
evaporation/fission). The SMM has been found to account successfully 
for a large variety of observables in heavy ion induced fragmentation 
reactions, including mass spectra and momentum distributions. We thus 
continue our previous fragmentation studies \cite{gait08} by including now 
as a new feature a coalescence scenario for the formation of (light) 
hypernuclei in dynamical transport simulations.

\section{Theoretical description of reactions}

The theoretical description of hadron-nucleus and heavy-ion reactions 
is based on the semiclassical kinetic theory of statistical mechanics 
\cite{kada}. The covariant extension of this equation
is the relativistic Boltzmann-Uehling-Uhlenbeck (RBUU) equation
\cite{blaettel}
\begin{eqnarray}
& & \left[
k^{*\mu} \partial_{\mu}^{x} + \left( k^{*}_{\nu} F^{\mu\nu}
+ M^{*} \partial_{x}^{\mu} M^{*}  \right)
\partial_{\mu}^{k^{*}}
\right] f(x,k^{*}) = \frac{1}{2(2\pi)^9} \nonumber\\
& & \times \int \frac{d^3 k_{2}}{E^{*}_{{\bf k}_{2}}}
             \frac{d^3 k_{3}}{E^{*}_{{\bf k}_{3}}}
             \frac{d^3 k_{4}}{E^{*}_{{\bf k}_{4}}} W(kk_2|k_3 k_4)
 \left[ f_3 f_4 \tilde{f}\tilde{f}_2 -f f_2 \tilde{f}_3\tilde{f}_4
\right] ~,
\label{rbuu}
\end{eqnarray}
for the $f(x,k^{*})$ $1$-body phase space distribution function for the 
hadrons. 
In particular, nucleons, hyperons and all resonances up to a mass of $2$ GeV, as 
well as mesons, e.g., pions,kaons, are explicitely propagated. 
In the collision term the short-hand notations $f_i \equiv f(x,k^{*}_i)$
for the particle and $\tilde{f}_i \equiv (1-f(x,k^{*}_i))$
for the hole distributions are used, with the $4$-momentum 
$k^{*\mu}=(E^{*}_{k},\vec{k})$ where 
$E^*_{\bf k} \equiv \sqrt{M^{*2}+{\bf k}^2}$.
The collision integral explicitly exhibits the final state Pauli-blocking while
the in-medium scattering amplitude includes the  Pauli-blocking of intermediate
states. 
The collision term includes 21 mesons and 60 resonances. We note that all elastic 
and inelastic processes, such as resonance 
production and absorption with associated meson production, are included. 
In particular, inelastic processes with strangeness production are explicitly 
accounted for, e.g., $BB\rightarrow BYK$, $B\pi\rightarrow YK$, $BK\rightarrow BK$ 
(with isospin exchange between $K^{0}$ and $K^{+}$), which are important for 
the production of hyperons. 
An exact solution of the set of the coupled transport equations 
for the different hadrons is not possible. Therefore, the commonly used 
test-particle method for the Vlasov part is applied, whereas the 
collision integral is modeled in a parallel-ensemble Monte-Carlo algorithm. 
The baryonic mean-field is modeled within the non-linear Walecka model 
(mean-field approximation of the QHD) \cite{QHD}. In the following calculations 
the $NL2$ parametrization of the non-linear Walecka model has 
been adopted, which gives reasonable values for the compression modulus 
($K=210$ MeV) and the saturation properties of nuclear matter 
\cite{blaettel}. For the parameters of the collision integral see 
Ref. \cite{larionov}. The numerical implementation for the solution of 
equation (\ref{rbuu}) is that of GiBUU \cite{GiBUU}.

The formation of hypernuclei depends on the way the fragmentation process 
between nucleons and, in particular, between nucleons and hyperons is 
accounted for in dynamical transport situations. It is well known, that 
the fragmentation process cannot be dynamically described within transport 
models of Boltzmann-type, since in such approaches only one-body 
phase-space densities are calculated and the dynamical phase-space 
evolution of physical fluctuations is neglected. In situations 
with long time scales for the fragmentation process, e.g. spectator 
dynamics in heavy-ion collision and dynamics in hadron-induced reactions, 
a statistical description of the fragmentation process is applicable, as 
has been shown in a previous work \cite{gait08}. 

The statistical multifragmentation model (SMM) \cite{botvina} has been 
widely applied to multiple fragment production. 
In its standard version \cite{botvina}, SMM does not account for 
strangeness degrees of freedom. Only 
recently it has been attempted to extend the statistical approach by 
including hyperons \cite{SMMH}, however, this work is still in 
progress. We thus apply the SMM model only for the statistical decay 
of the excited nucleonic spectators. The formation of spectator 
hypernuclei is then modeled through a phase space coalescence between 
the fragments and the hyperons originating in the fireball. If hyperons are inside the 
(fragmenting) spectators and their velocities coincide with those of 
the fragments, they are supposed to form a hyperfragment. 
This is the basic idea of the coalescence model \cite{coala}, which 
has been applied in the past to describing hypernuclei, see Refs.
\cite{Wakai1,Wakai2}. In those models one assumes that the hypernuclear 
production cross section factorizes into the cross section for 
hyperon production and that for the fragment production. Phase space 
coalescence effects are accounted for by a coalescence function, which 
enters into the hypernucleus cross section and can be determined within 
the density matrix formalism, as described in detail in Ref. \cite{Wakai1}. 
From those studies the coalescence velocity $R$ can be 
also extracted. $R$ defines the radius in the 
velocity space, inside of which particles can form a (hyper)cluster. From 
those studies it turned out that the coalescence velocity 
is less or in the order of the Fermi-velocity, which agrees with 
phenomenological investigations of Ref. \cite{WakaiRest}.

In this work, the coalescence velocity $R$ for the formation of 
hyperfragments is a phenomenological parameter, which should, 
in principle, be adjusted to empirical information. However, since there 
are no precise data on hypernuclear formation available yet, we have chosen 
$R$ such as to produce results as close as possible to 
predictions of the coalescence models of Refs. \cite{Wakai2}.

We have applied the transport model together with subsequent SMM and 
phase space coalescence calculations to ${}^{12}C+{}^{12}C@2~AGeV$ 
collisions in spectator fragmentation and to $p+{}^{12}C@50~GeV$ 
reactions. The reason for the choice of these colliding systems is 
that they will be experimentally investigated by the HypHI-collaboration 
at GSI and JPARC facilities, respectively. We thus present here for the 
first time estimates for such reactions from a dynamical model. 
However, before presenting the results on hypernuclei, a benchmark study 
is first discussed in the next section.

\section{Benchmarks}

\begin{figure}[t]
\unitlength1cm
\begin{picture}(8.,7.5)
\put(2.0,0.0){\makebox{\epsfig{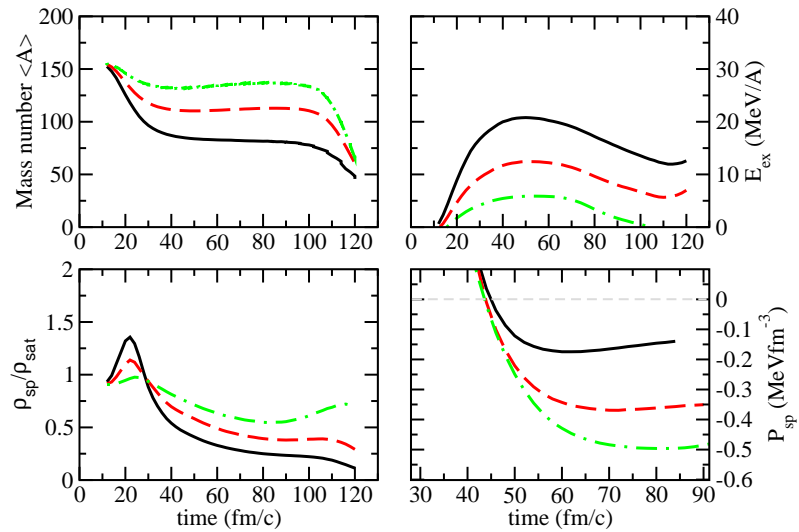}}}
\end{picture}
\caption{Time dependence of the average mass number (top-left panel), 
the excitation energy per nucleon (top-right panel), the local baryon 
density (bottom-left panel) and the local pressure (bottom-right panel) of 
projectile spectators in Au+Au collisions at $0.4~AGeV$ incident energy. The 
different curves at each panel indicate calculations at different impact 
parameters $b=6,~8,~10~fm$ (solid, dashed, dot-dashed, respectively).
}
\label{Fig1}
\end{figure}
According to the coalescence scenario \cite{Wakai1} the production 
cross section of hypernuclei depends mainly on $3$ parameters, 
the cross sections for strangeness and fragment production 
and the coalescence velocity. In this section we test 
the first two parameters from transport calculations and discuss 
the time evolution of spectator properties.

The application of the combined GiBUU+SMM approach has been found to work 
very well in proton induced reactions in terms of cross sections of 
global mass and charge distributions and, in particular, in terms of 
neutron number distributions of a wide selection of different isotopes 
\cite{gait08}. However, in heavy-ion collisions (HIC) the situation 
is more complex with respect to that in proton-induced reactions. 
Depending on the centrality one has to clearly 
separate strongly interacting participant matter ("fireball") from less 
dynamically developed one ("spectators"). In a theoretical study this can 
be easily done just from the knowledge of suffered collisions for each particle. 
However, the experimental situation is less trivial, and several phase space 
limitations have to be imposed to separate spectators from fireball sources as 
precisely as possible. In the ALADIN experiment, which we will discuss 
in this section, a rapidity selection has been performed \cite{ALADIN}. 
Particles with $y \ge 0.75 y_{proj}$ are assumed to belong to the spectator 
sources. In the theoretical calculations we have performed the same selection 
for consistency. In addition a density cut of $\rho>\rho_{sat}/300$ 
(with $\rho_{sat}=0.16~fm^{-3}$ being the nuclear saturation density) has 
been applied in order to separate particles bound to spectators from emitted 
ones.

\begin{figure}[t]
\unitlength1cm
\begin{picture}(10.,7.0)
\put(2.5,0.){\makebox{\psfig{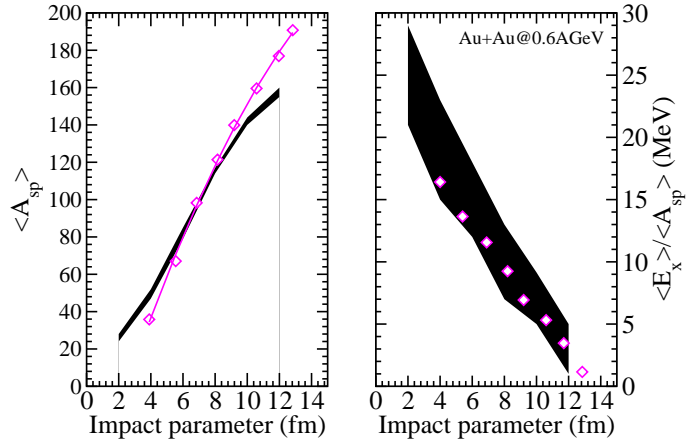}}}
\end{picture}
\caption{Impact parameter dependence of the average mass number (left panel) 
and the excitation energy per nucleon (right panel) for projectile-spectator 
matter. Theoretical calculations (black band) are 
compared with data from the ALADIN collaboration \protect\cite{ALADIN} 
(open diamonds).
}
\label{Fig2}
\end{figure}
Fig.~\ref{Fig1} shows the time evolution of the projectile-spectator 
in terms of its mass number, excitation energy and pressure and for particular 
centralities. The density $\rho_{sp}$ and pressure $P_{sp}$ have been 
extracted at the densest point of the spectator region. After its 
formation, the projectile source stabilizes after the freeze-out time of 
$t_{fr}\sim 50-70~fm/c$ as indicated by constant values 
of mass number and excitation energy. Furthermore, at 
$t\sim t_{fr}$ the local pressure \cite{Press} becomes negative and 
almost constant with time. 
Also the slope $dP_{sp}/d\rho_{sp}$ becomes negative after 
$t>55, 65, 75~fm/c$ for $b=6,8,10~fm$, respectively. This feature may indicate 
the onset of spinodal instabilities, 
assuming a constant 
temperature evolution in the center of the spectator. However, fragment 
formation cannot be described in transport theory. 
We, therefore, proceed as follows: The spectator matter 
reaches local equilibrium at $t\approx 40~fm/c$, corresponding to momentum space 
isotropy, i.e., equality of longitudinal and transverse pressure components. Hence 
the dynamical evolution has come to an end. This slightly excited spectator 
configuration is then taken as the initial state for fragmentation which is treated 
as a statistical process according to the laws of thermodynamics as realized in the 
SMM approach. The physics discussed here is similar also for higher energies 
(see next section) and has been recently applied in describing spallation 
reactions \cite{gait08}.

\begin{figure}[t]
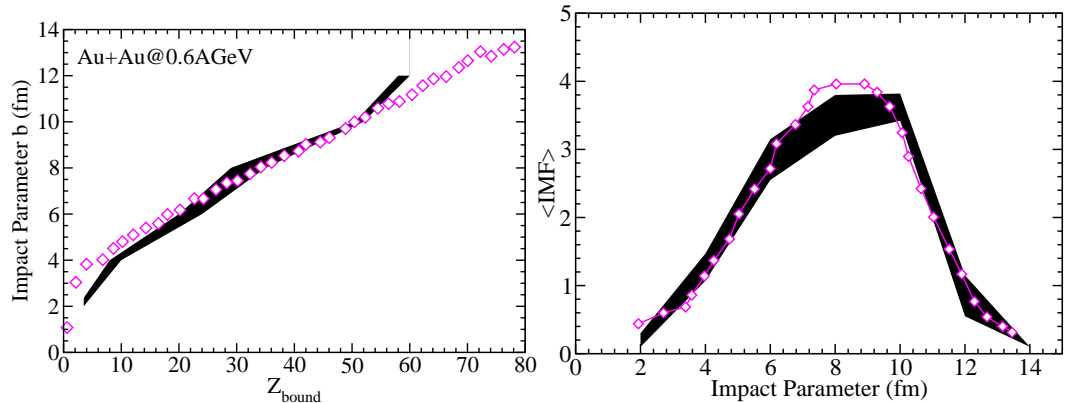

\unitlength1cm
\begin{picture}(10.,6.0)
\put(0.0,0.){\makebox{\psfig{file=Fig3a.eps,width=7.0cm}}}
\put(7.0,0.){\makebox{\psfig{file=Fig3b.eps,width=7.0cm}}}
\end{picture}
\caption{Left panel: Relation  between the impact parameter and the 
observable $Z_{bound}$. Right panel: Impact parameter dependence of the 
mean multiplicity of Intermediate-Mass-Fragments (IMF). Theoretical 
calculations (black band) are compared with data from the ALADIN 
collaboration \protect\cite{ALADIN} (open diamonds).
}
\label{Fig3}
\end{figure}
The calculated parameters of the spectator source are shown in Fig.~\ref{Fig2} and compared 
with data from the ALADIN collaboration \cite{ALADIN}. The black bands in the 
theoretical calculations indicate the effects of moderate changes on mass and 
excitation energy at the time near the onset of equilibration. 
The comparison with the data is reasonable except for the very peripheral events, 
in which the average mass of spectator sources is theoretically 
underestimated. In this respect the ground state description for the 
initial nuclei within the GiBUU approach (or within any other transport model) 
becomes important. If the initial nuclear configuration is not a solution of the 
Hamiltonian underlying the transport model the early stage of the dynamical 
evolution include artificial density oscillations which may lead to spurious 
particle emission processes. Results for very peripheral reactions may be thus 
affected to some extent. An improved initialization method in the spirit 
of a covariant density functional approach is under study \cite{ExRTF}.

Applying these parameters of the spectator source to the SMM model, we arrive at the 
results on spectator fragmentation, see Figs.~\ref{Fig3}. Here the relation between 
the impact parameter and the observable $Z_{bound}=\sum_{Z\ge 2} Z$ (with $Z$ being the charge) 
and the average number of intermediate mass fragments ($3 \le Z \le 30$) as function 
of the impact parameter are displayed. The observable $Z_{bound}$ is well 
reproduced for all centralities, but less so for the most peripheral ones (see 
discussion above). Interesting is the reasonable description of the $IMF$, 
even for peripheral collisions. This is an indication of stronger 
evaporation effects in the theoretical calculations; the underestimate in the mass 
distribution is compensated by the slightly higher values of the excitation 
energy in the theoretical calculations, see again Fig.~\ref{Fig2}.

\begin{figure}[t]
\unitlength1cm
\begin{picture}(10.,6.0)
\put(1.75,0.){\makebox{\psfig{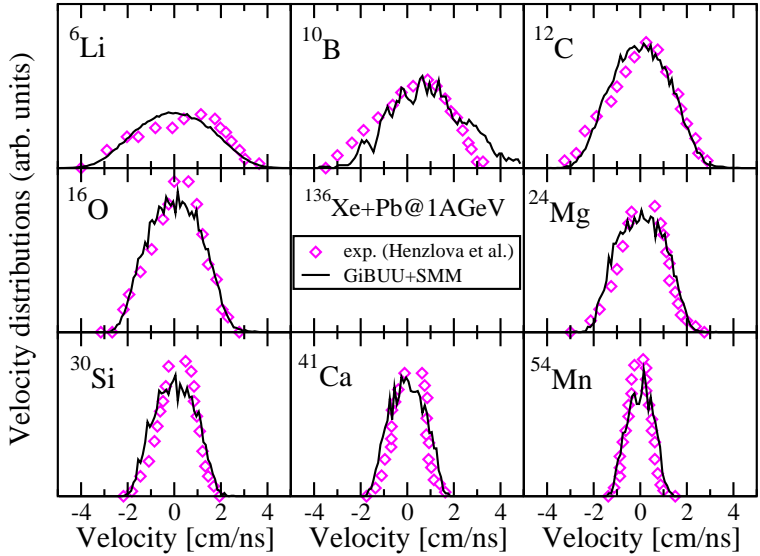}}}
\end{picture}
\caption{Longitudinal velocity distributions in the projectile 
frame for ${}^{136}Xe+Pb@1~AGeV$ reactions. Theoretical calculations 
(solid curves) are compared with experimental data (open diamonds) from 
\protect\cite{XePb}.
}
\label{Fig3c}
\end{figure}
So far it has been demonstrated that inclusive absolute yields are reproduced well 
in the hybrid GiBUU+SMM approach. However, a correct description of velocity 
distributions of statistically produced fragments is also crucial for their subsequent 
coalescence with hyperons. Fig. ~\ref{Fig3c} shows fragment velocity 
distributions in the projectile frame in comparison with experimental data taken 
from \cite{XePb}. The velocity distributions are described rather well on a 
quantitative level, in particular the width of the Gaussian-like fragment velocity 
distributions is well reproduced.

Similar studies on multifragmentation were performed 
within nuclear molecular dynamics and intranuclear cascade models. More 
extensive reviews can be found in Refs. \cite{NMD} and \cite{INC}. 
Furthermore, we note that in the past similar hybrid approaches have been applied in 
dynamical situations. It turned out that momentum distributions are reproduced 
fairly well in heavy-ion collisions as well as in hadron- and pion-induced 
reactions, see, e.g., Ref. \cite{botvina}, which is an important issue before 
applying coalescence for the formation of hypernuclei.

\begin{figure}[t]
\unitlength1cm
\begin{picture}(10.,6.0)
\put(1.75,0.){\makebox{\psfig{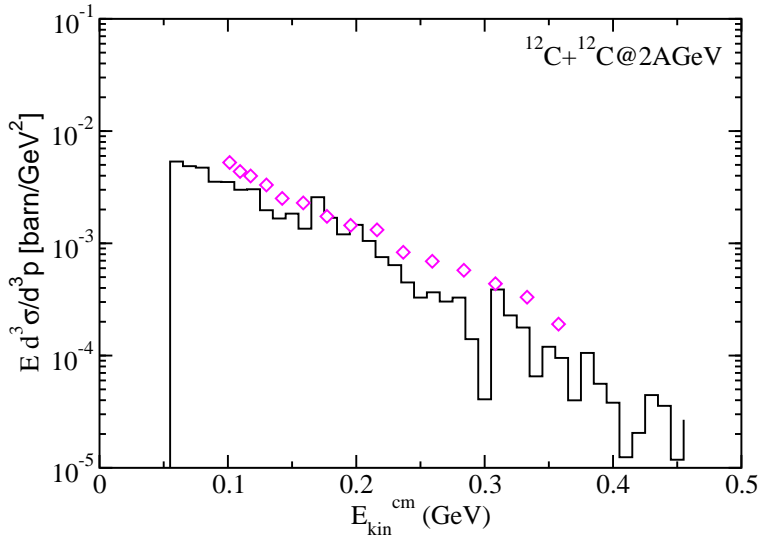}}}
\end{picture}
\caption{Inclusive invariant production cross section of $K^{+}$ mesons 
in $C+C@2AGeV$ collisions at a polar angle of $\Theta_{lab}=40^{o}$. 
Theoretical calculations (histogram) are compared with data from 
the KaoS collaboration  \protect\cite{KaoS} (open diamonds).
}
\label{Fig4}
\end{figure}

For the formation of hypernuclei the production and propagation of 
strangeness degrees of freedom play also an important role. We have 
tested, that for the light $C+C$-system, which will be studied later, 
the production cross sections of $K^{+}$ mesons are well reproduced 
(see Fig.~\ref{Fig4}), although the relativistic mean-field model 
gives a rather strong momentum dependence at these high energies. 
A more detailed discussion of strangeness production can be found in 
Ref. \cite{larionov,KaonsLarionov}. 

We conclude that the combined GiBUU+SMM model describes inclusive 
multifragmentation observables at moderate relativistic energies and also 
strangeness production very well. Since strangeness production is also reproduced 
it is interesting to extend this model it to the description of hypernuclear 
formation in high energy heavy-ion collisions and proton-induced reactions. These 
topics will be discussed in the next two sections.

\section{Hypernuclei Formation in heavy-ion collisions}

\begin{table}[t]
\begin{center}
\caption{\label{table1} \small{
Inclusive production cross sections for different types of hypernuclei (as 
indicated) for the colliding system ${}^{12}C+{}^{12}C@2~AGeV$. The 
contribution from exclusive pion-nucleon scattering, i.e. 
$\pi N \rightarrow YK$, is shown separately.
}}
\vskip 0.5cm
\begin{tabular}{|l|c|c|c|c|}
\hline\hline 
 & ${}^{4}_{\Lambda}H$  & ${}^{4}_{\Lambda}He$ & ${}^{5}_{\Lambda}He$ \\ 
\hline\hline
   total yield ($\mu b$)                 & 2.2 & 4 & 1.4 \\ 
\hline
   pionic contribution ($\mu b$)   & 0.3  & 0.2 & 0.03  \\ 
\hline\hline
\end{tabular}
\end{center}
\vskip 0.5cm
\vskip 0.5cm
\end{table}

The production of hypernuclei in energetic collisions between light 
nuclei is one of the major projects being investigated by the 
HypHI-collaboration at GSI. 
The reason for selecting very light systems is the easier identification 
of hypernuclei via the weak decay of the hyperon into pions. In earlier 
theoretical studies, see, e.g., Refs. \cite{Wakai1,Wakai2,WakaiRest}, 
cross sections of the order of only a few microbarn ($\mu b$) were predicted, 
which can be easily understood: in collisions between very light systems, such 
as ${}^{12}C+{}^{12}C@2~AGeV$, secondary re-scattering effects inside the 
spectator matter, important for producing low energetic hyperons, 
are rare processes due to the small interaction volume. 
The situation is different in collisions between heavy nuclei 
due to the high production rate of strangeness and many secondary scattering 
events.

\begin{figure}[t]
\unitlength1cm
\begin{picture}(10.,8.0)
\put(0.0,0.){\makebox{\psfig{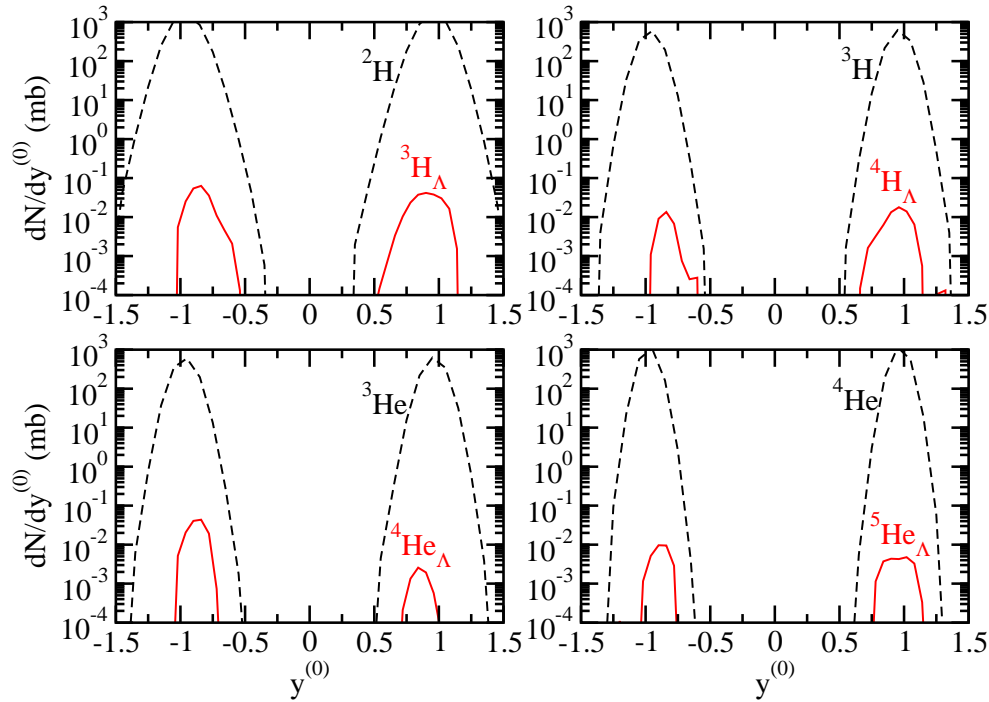}}}
\end{picture}
\caption{Rapidity distributions as function of the rapidity 
$y^{(0)}$ (normalized to the projectile rapidity in the c.m. frame) of 
different particle types (as indicated) for the system 
${}^{12}C+{}^{12}C@2~AGeV$. 
}
\label{Fig5}
\end{figure}

The coalescence prescription (in coordinate and momentum space, see again 
section $2$) for the formation of hypernuclei has been 
applied to the description of spectator fragmentation in $C+C@2AGeV$ 
collisions. The separation between spectator and participant matter 
has been adjusted by a rapidity criterion of $|y| > 0.7y_{proj,targ}$. 
Inclusive rapidity spectra for different light fragments and 
hyperfragments from spectator matter are shown in Fig.~\ref{Fig5}. 
The estimated hyperfragment production is  $\approx 5-6$ orders of magnitude 
less than that of fragment production in general. This is obvious, since the 
strangeness production cross sections from exclusive primary channels, i.e. 
primary $BB\rightarrow BYK$, $BB\rightarrow BBK\overline{K}$, and 
secondary ones ($B\pi\rightarrow YK$ and 
$B\overline{K}\rightarrow \pi Y$, $B$ stands for a baryon and $Y$ for a 
hyperon) are very low (orders of few $\mu b$) 
\cite{tsushima}. Among the different processes contributing to the 
formation of hypernuclei, $BB\rightarrow BYK$ and 
$\Lambda B\rightarrow \Lambda B$ and the secondary one 
$B\pi\rightarrow YK$ give the major contribution to the formation of 
hypernuclei. Strangeness production channels with $3,4$-body final states 
are important in order to produce low energy hyperons that can be 
captured by spectator matter. The same argument also holds for secondary 
re-scattering via elastic hyperon-nucleon and pion-nucleon processes.

The transport results are summarized in table~\ref{table1}, in which 
the total inclusive hypernuclei production cross sections 
(first line) and the contributions originating from 
pion-nucleon scattering (inside the spectator matter) are shown. 
The transport calculations predict only moderate contributions to 
the total hyperfragment cross section from indirect coalescence through 
the $\pi B$-channel. We have investigated this point in more detail by 
applying the SMM model to different freeze-out times (not shown here) with 
the result of a rather early formation of hypernuclei. Thus, 
we conclude that hypernuclei are mainly formed due to the capture of 
fireball hyperons during the passage stage of the spectators near the 
expanding fireball region, with subsequent rescattering of hyperons with 
spectator matter, i.e. $\Lambda B\rightarrow \Lambda B$. This results was 
also confirmed by earlier model calculations from Wakai et al. \cite{Wakai2}, 
in which at low relativistic energies around $2~AGeV$ the coalescence between 
fragments and hyperons from $\pi B$ collisions is only of minor importance, 
which is obvious for collisions of light systems at low energies. 
In principle, antikaons $\bar{K}$ may also contribute to hypernuclear 
production. However, the formation of hypernuclei through antikaons has not 
been considered here due to their very low production cross sections with 
very high threshold value , e.g. $BB\rightarrow BBK\overline{K}$. 

We conclude from our exploratory calculations, that hypernucleus formation in 
$C+C@2AGeV$ reactions takes place with estimated 
production probabilities on a few $\mu b$. To substantiate this result further 
theoretical extensions would be welcome. For example, a purely statistical 
treatment of excited fireball-like spectators including strangeness content seems 
possible \cite{SMMH}.

\section{Hypernuclear Formation in high energy $p+{}^{12}C$ reactions}
\begin{table}[t]
\begin{center}
\caption{\label{table2} \small{
Inclusive production cross sections for different types of hypernuclei (as 
indicated) for the system $p+{}^{12}C@50~GeV$. The 
contribution from pion-nucleon scattering, i.e. $\pi N \rightarrow YK$, is shown 
separately.
}}
\vskip 0.5cm
\begin{tabular}{|c|c|c|c|c|c|c|c|}
\hline\hline 
 & $\sigma_{tot}~[\mu b]$ & $\sigma_{pionic}~[\mu b]$ & & $\sigma_{tot}~[\mu b]$ &
$\sigma_{pionic}~[\mu b]$ \\ 
\hline\hline
${}^{3}_{\Lambda}H$ & 43 & 33 & ${}^{3}_{\Lambda}He$ & 0.9 & 0.8 \\
\hline
${}^{4}_{\Lambda}H$ & 28 & 14 & ${}^{4}_{\Lambda}He$ & 35 & 16 \\
\hline
${}^{5}_{\Lambda}H$ & 9.5 & 2.6 & ${}^{5}_{\Lambda}He$ & 9.8 & 4.3 \\
\hline
${}^{6}_{\Lambda}H$ & 0.8 & 0.2 & ${}^{6}_{\Lambda}He$ & 16 & 1.3 \\
\hline\hline
\end{tabular}
\end{center}
\vskip 0.5cm
\vskip 0.5cm
\end{table}

The colliding system $p+{}^{12}C@50GeV$ will be investigated 
at the new J-PARC facility in Japan \cite{JPARC}. As in the 
case of low-energy proton induced reactions \cite{gait08} and 
high energy heavy ion collisions (see previous section), we apply 
also here the same dynamical approach for the pre-equilibrium 
stage. We note that the string fragmentation model in terms of 
the PYTHIA high energy packages \cite{PYTHIA} is included in 
the transport simulations.

As an interesting feature here, transport simulations 
show a dynamical break-up of the excited residual nucleus 
into a remnant in the target rest frame and a moving source. 
Due to the small interaction region and thus small effect 
of re-scattering inside the initial compound system, the target 
nucleons which collide with the beam protons maintain their high 
momenta after the collision. This finally leads to a 
pre-equilibrium break up of the initial compound system into 
a moving source, which contains highly excited baryon 
and also anti-baryon states. 
\begin{figure}[t]
\unitlength1cm
\begin{picture}(10.,8.0)
\put(0.,0.){\makebox{\psfig{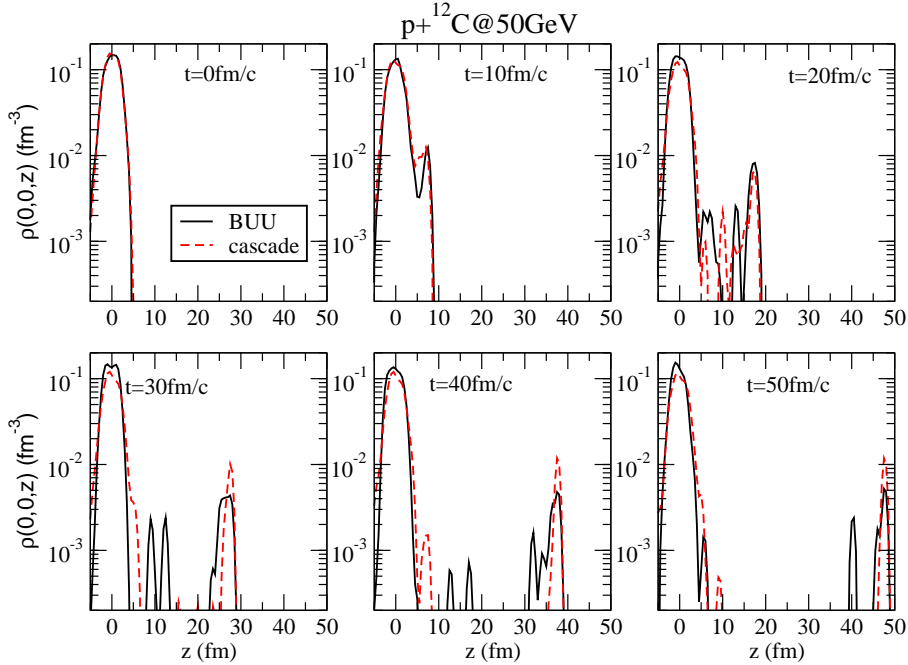}}}
\end{picture}
\caption{Density profiles $\rho(0,0,z)$ along the beam axis 
at different time steps as indicated. Calculations with (solid) 
and without mean-field potential (dashed) are shown. The considered system 
is $p+{}^{12}C@50GeV$ at a fixed impact parameter of $b=2.5~fm$.
}
\label{Fig6a}
\end{figure}
This feature is shown in Fig.~\ref{Fig6a} in terms of the temporal 
evolution of the density profiles along the $z$-axis. This 
particular situation appears even in calculations without 
using any mean-field potential (cascade mode), as one would 
expect from moderate mean-field effects on the reaction dynamics 
at so high energies. This is not obvious due to the strong momentum 
dependence of the underlying relativistic mean-field model 
which linearly rises with energy, a well known feature 
of mean-field models of Walecka type \cite{MDI}. In particular, 
the missing momentum dependence in the cascade calculations leads 
to a moderate density enhancement of the moving source due to the 
missing repulsion of the momentum dependent part of the mean-field. 
However, one should note this uncertainty of the momentum dependence 
of the mean-field at so high energies, in which there is no experimental 
information available, only moderately affects the yields of produced 
particles strange baryons and pions, in particular the differences 
between the cascade calculations and the full BUU calculations differ 
only by less than $5\%$. By considering inclusive reactions integrated 
over the entire centrality region the momentum dependent effect 
decreases. Therefore we do not expect considerable effects on the 
production of hypernuclei due to the uncertainty in the momentum 
dependence of the mean-field and continue the discussion on hypernuclei, 
which is the main topic of this work.

A clear separation between processes relevant for the formation 
of hypernuclei and others is a non-trivial task in 
reactions at very high relativistic energies. However, it is 
possible to give a rough picture using qualitative arguments. 
Relevant for the formation of hypernuclei are low energy hyperons, 
as in the case of heavy-ion collisions discussed in the previous 
section. Here processes with hyperon production in many-body 
final states are crucial as well as re-scattering of initially 
produced pions with particles of the moving source. In contrast 
to the situation of low-energy heavy-ion collisions the produced 
pion yields are now very high, i.e. a few $b$ (compared to the 
yields in the $mb$ region for other particles). 

Concerning the fragmentation mechanism the situation is similar to that of 
the previous section. As described 
in section $2$, non-equilibrium dynamics is responsible for a de-excitation 
in terms of binary processes with multi-meson production, e.g., pions, and 
the achievement of thermal equilibrium for both systems, the remnant nuclei  
at rest and the moving source. We have checked in terms of the anisotropy 
ratio \cite{gait08} that both systems approach thermal equilibrium 
at $t>(15-20)~fm/c$ (depending on centrality). Afterwards the SMM has been 
separately applied to fragmentation of both sources, i.e. the residual 
nuclei at rest and the moving source. In the later case the SMM model 
has been applied to moving sources consisting of 4 and more particles, 
otherwise a simple coalescence (in coordinate and momentum space) has been 
adopted.

\begin{figure}[t]
\unitlength1cm
\begin{picture}(10.,8.0)
\put(0.,0.){\makebox{\psfig{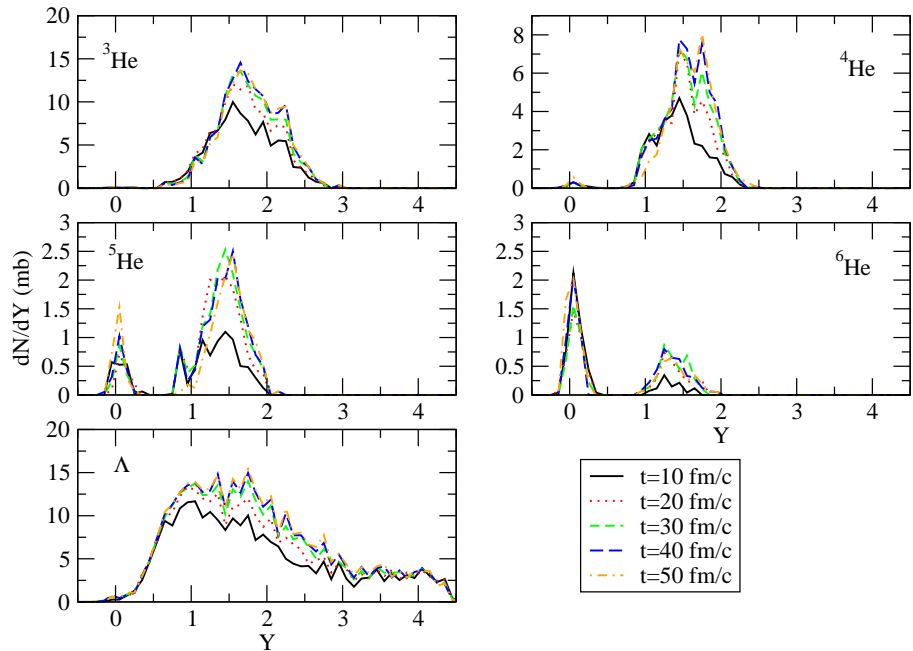}}}
\end{picture}
\caption{Rapidity distribution of different particle types (light 
fragments and hyperons) at different time steps, as indicated, for 
the system $p+{}^{12}C@50GeV$.
}
\label{Fig6}
\end{figure}
The whole situation is illustrated in Fig.~\ref{Fig6} in terms of the 
temporal evolution of the rapidity distributions of different particle 
types (fragments and hyperons), as indicated. First, a 
$2$-component structure of the source can be clearly seen 
for the heaviest fragments. The rapidity spectra approach 
a freeze-out state after $t>30-40~fm/c$, in which the nucleonic content 
of the moving source is enhanced due to, e.g., resonance decays. The 
time behavior of the $\Lambda$-rapidity distribution demonstrates the 
features discussed above. In particular, some of the hyperons are produced 
inside the moving sources in the same rapidity region as that of the 
fragments. The role of initially produced pions will be also of relevance 
for the hypernucleus formation (see below). These effects indicate 
a production of hypernuclei inside the moving sources.

\begin{figure}[t]
\unitlength1cm
\begin{picture}(10.,8.0)
\put(0.,0.){\makebox{\psfig{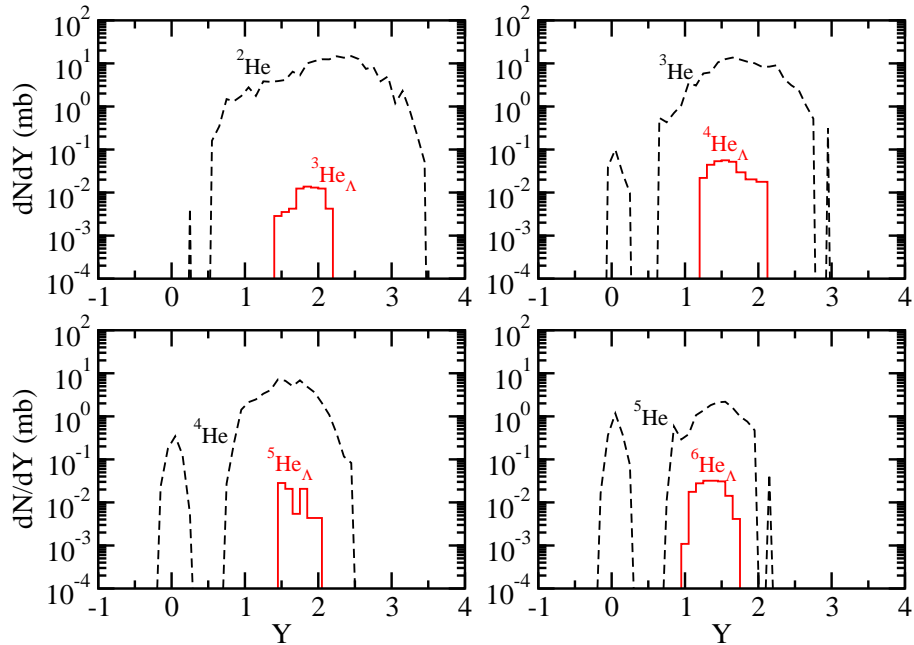}}}
\end{picture}
\caption{Rapidity distributions of different particle types for the 
system $p+{}^{12}C@50GeV$. 
}
\label{Fig7}
\end{figure}

Fig.~\ref{Fig7} shows the inclusive rapidity spectra 
for $He$-isotopes, and the corresponding hypernuclei, 
respectively. The formation of hypernuclei takes mainly place 
inside the moving sources with energies per nucleon in 
the range of view $GeV$. Due to extremely high multiplicities of mesons, 
in particular pions, the pion-nucleon scattering contribution 
to the formation of hypernuclei is now the dominant process for the 
lightest hypernuclei, which it was not the case at the lower energies 
of the previous section. This result, which is again in line with earlier 
theoretical investigations \cite{Wakai2}, is demonstrated in table 
\ref{table2}  for the total yields of different hypernuclei (first and third columns) 
and those yields originate only from pion-nucleon scattering. Thus, 
meson-nucleon re-scattering should play an important role in the 
formation and propagation of hypernuclei in highly energetic 
proton-induced reactions. Other processes of similar type may also be 
of relevance, e.g., $\overline{K}N \rightarrow Y\pi$.

\section{Conclusions and outlook}

We have theoretically investigated the different mechanisms for 
hypernuclear production in reactions relevant to future experiments 
in hypernuclear physics. We have applied an extended GiBUU+SMM transport 
approach which accounts for the dynamical pre-equilibrium phase space 
evolution and also for the statistical decay of the asymptotically 
equilibrated spectator sources leading to spectator fragmentation. The 
formation of hyperfragments has been modeled within a phase space 
coalescence model.

The reasonable description of spectator fragmentation at intermediate energies 
and the fragmentation of residual nuclei in proton-induced reactions has 
served as the starting basis to extend the transport studies by 
accounting for fragments with strangeness degrees of freedom (hypernuclei). 

We have applied the model to heavy-ion collisions at intermediate relativistic 
energies and to proton-induced reactions at very high incident energies by 
analyzing the transport calculations in terms of formation of light 
hypernuclei. We have estimated the production cross sections of light 
hypernuclei in such reactions by exploratory transport theoretical calculations. 

This first transport theoretical study on hypernuclear production in 
reactions at low and high incident energies can be further extended. 
In particular, a future combination of our transport model with the recently 
developed statistical multifragmentation model 
including strangeness degrees of freedom would be a helpful tool. Furthermore, 
a detailed study on possible influences of the hadronic mean-field 
on hypernuclear formation may be important. Within a relativistic framework 
one can extend the present non-linear Walecka model to more general hadronic 
matter including strangeness degrees of freedom \cite{lenske}, which has been 
partially implemented and is being tested.

In summary, transport simulations predict significant production 
cross sections for highly energetic hypernuclei. The theoretical 
estimates may be useful for the future experiments at GSI and the 
J-PARC facility, since high energy 
hypernuclei can be easily separated from the background and 
detected. The production of hypernuclei with strangeness $S=-2$ may 
be also of major importance, e.g., $\Xi$- and 
double-$\Lambda$-hypernuclei. However, due to very low statistics 
these processes have not been studied in this Letter. 
We conclude that this work provides an appropriate theoretical basis 
for investigations on hypernuclear physics.

{\it Acknowledgments.} We would like to thank for many useful discussions 
A. Botvina. A.B. Larionov and I.N. Mishustin. We thank T. Saito and 
the members of the HypHI Collaboration for discussions on their proposal 
to study the colliding systems. Finally, we would like 
to thank the GiBUU group for useful discussions. This work is 
supported by BMBF.



\begin{thebibliography}{99}
\bibitem{Press}
The local pressure is extracted as the average value of 
the diagonal components of the energy-momentum tensor, 
calculated in the rest frame of the spectator sources.
\end{thebibliography}
\end{document}